# Spin current and electrical polarization in GaN double-barrier structures.

V.I. Litvinov

*WaveBand/ Sierra Nevada Corporation,*

*15245 Alton Parkway, Suite 100*

*Irvine, CA 92618*

**Abstract.**

Tunnel spin polarization in a piezoelectric AlGaN/GaN double barrier structure is calculated. It is shown that the piezoelectric field and the spontaneous electrical polarization increase an efficiency of the tunnel spin injection. The relation between the electrical polarization and the spin orientation allows engineering a zero magnetic field spin injection manipulating the lattice-mismatch strain with an Al-content in the barriers.

Spin-dependent electron tunneling in III-V semiconductors has become a subject of an intensive study in regards with a search for an effective non-magnetic spin injector. Non-magnetic spin filters are appealing as they do not require a magnetic field and allow spin manipulation to be achieved with an applied voltage.

It is known that the direct contact between a ferromagnetic metal and a semiconductor has low spin injection efficiency due to the conductivity mismatch [1]. To improve an efficiency of spin transfer between a metal and a semiconductor, electrical contacts of enlarged resistance have been proposed in Ref.[2]. Possible implementation of this approach is tunnel contacts between nonmagnetic semiconductors with an intrinsic spin splitting. Spin selection in tunnel contacts occurs in the course of electron transmission through a single- or double-barrier- quantum well (QW) providing the spin-orbit interaction is essential. Two sources of the momentum-dependent spin-splitting, an interface-induced Rashba interaction and bulk Dresselhaus terms were considered to date exclusively in InGaAs-based zinc-blende (ZB) spin resonant tunneling diodes (SRTD) [3-5].

This paper deals with the voltage-controlled tunneling spin injection in wide bandgap semiconductors. Estimated theoretically [6,7] and found experimentally [8,9] the spin-splitting in Al(In)GaN/GaN QW have a magnitude comparable to that of their GaAs counterparts. This suggests that the wide bandgap SRTD could benefit those spintronic applications that require an efficient spin injector. Recent experimental data on InGaN superlattices show that the zero magnetic field spin polarization depends on an internal strain [10], which is inevitable in GaN-based devices due to a lattice mismatch between layers and a substrate.

In this paper we study the spin tunneling and the role the strain plays in spin injection in a wurtzite wide bandgap SRTD. It is useful to note the difference between the two types of QWs: InGaAs-based (001)-oriented ZB and GaN-based (0001)-wurtzite (W). The voltage applied across the (001)-oriented structure induces the electric fields of the same sign in each layer and the flat-band approximation well describes the electron band edges if the external bias turns zero. In a W-QW the built-in electric fields, caused by spontaneous and lattice-mismatch piezoelectric polarizations, distort the band profile causing the Zener tunneling to occur even if the external bias is not applied. It is shown that the built-in fields make a difference as for the voltage-controlled spin tunneling, namely, the spontaneous and strain-induced electrical polarization fields increase spin polarization efficiency that otherwise would be low if no electrical polarization were taken into account.

The electron Hamiltonian in each (0001)||z-oriented layer includes the kinetic energy and the spin-orbit interaction as follows[6]:

$$H = \frac{p_\parallel^2}{2m_\parallel} + \gamma_t \left(\sigma_x k_y - \sigma_y k_x\right) k_\parallel^2 + k_z \left[\frac{\hbar^2}{2m_z} + \gamma_l \left(\sigma_x k_y - \sigma_y k_x\right)\right] k_z + \left(\lambda + \frac{\partial \beta}{\partial z}\right)\left(\sigma_x k_y - \sigma_y k_x\right) + V(z),$$  (1)

where $\frac{\partial \beta}{\partial z}$ is the Rashba coupling coefficient in GaN, $\lambda$ is the bulk linear spin-orbit coupling constant, $\gamma_{t,l}$ are the Dresselhaus-type interaction constants, respectively, $m_\parallel, m_z$ are the effective masses, $p_z = -i\hbar \frac{\partial}{\partial z}$ is the electron momentum in the growth direction, $\vec{\sigma}$ are the Pauli matrices, $V(z)$ is the heterostructure potential energy that accounts for an external bias $V_{ext}$. The coupling constant $\gamma_t$ in Eq.(1) renormalizes the in-plane effective mass and will be neglected throughout the paper.

After the unitary transformation from the initial $z$-oriented spinors to a new basis $|u_\pm> = \frac{1}{\sqrt{2}}\begin{pmatrix}1 \\ \pm ie^{i\varphi}\end{pmatrix}$, the Hamiltonian becomes diagonal as written below:

$$H_\pm = \frac{p_\parallel^2}{2m_\parallel} + k_z \frac{\hbar^2}{2m_z^\pm} k_z \mp \left(\lambda + \frac{\partial \beta}{\partial z}\right) k_\parallel + V(z), \quad \tan\varphi = k_y/k_x,$$

$$m_z^\pm = m_z \left(1 \mp \frac{2\gamma_l m_z k_\parallel}{\hbar^2}\right)^{-1}. \qquad (2)$$

Charge and spin currents will be calculated in the tunneling structure comprising five regions: left and right thick GaN leads (regions 1 and 5 in Fig.1) and the GaN QW placed in between two AlGaN barriers. Since the whole structure is lattice-matched to GaN, we account for the tensile strain in the barriers. Corresponding conduction band offsets, polarization fields (spontaneous and piezoelectric), and total internal electric fields $F_j$ in each layer of thickness $d_j$, were calculated as described in Ref. [11]. An example of the resulting conduction band profile is shown in Fig.1.

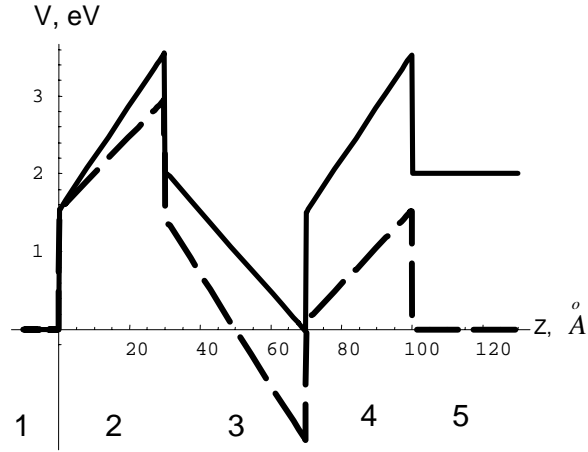

Fig.1. Conduction band profile: $30 \overset{\circ}{A}/ 40 \overset{\circ}{A}/ 30 \overset{\circ}{A} - Al_x Ga_{1-x} N / GaN / Al_y Ga_{1-y} N$ ($x = y = 1$), continuous line: $V_{ext} = -2V$, broken line: $V_{ext} = 0$.

Electron wave functions in layers, numbered in Fig.1, can be represented as follows:

$$\Phi_{j\pm} = u_\pm \Psi_{j\pm}(z) \exp\left[i\vec{k}_\parallel \vec{r}_\parallel\right],$$

$$\Psi_{1\pm} = e^{ik_1 z} + r_\pm e^{-ik_1 z}, \quad \Psi_{2\pm} = A_{2\pm} Ai(\rho_{2\pm}) + B_{2\pm} Bi(\rho_{2\pm}), \qquad (3)$$

$$\Psi_{3\pm} = A_{3\pm} Ai(\rho_{3\pm}) + B_{3\pm} Bi(\rho_{3\pm}), \Psi_{4\pm} = A_{4\pm} Ai(\rho_{4\pm}) + B_{4\pm} Bi(\rho_{4\pm}),$$

$$\Psi_{5\pm} = t_\pm e^{ik_5(z-d_2-d_3-d_4)},$$

where $Ai(s), Bi(s)$ are the Airy functions, $r_\pm, t_\pm$ are the reflection and transmission amplitudes, respectively,

$$\hbar k_1 = \sqrt{2m_z\left(E - \frac{\hbar^2 k_\parallel^2}{2m_\parallel}\right)}, \hbar k_5 = \sqrt{2m_z\left(E - \frac{\hbar^2 k_\parallel^2}{2m_\parallel}\right) + 2qm_z V_{ext}},$$

$$\rho_{2\pm} = C_{2\pm}\left(qF_2 z + \Delta E_{c1} - E\right), \rho_{3\pm} = C_{3\pm}\left[qF_3(z-d_2) + qF_2 d_2 - E\right],$$

$$\rho_{4\pm} = C_{4\pm}\left[qF_4(z-d_2-d_3) + qF_3 d_3 + qF_2 d_2 + \Delta E_{c2} - E\right], \qquad (4)$$

$$C_{j\pm} = \left(\frac{2m_{jz}^\pm}{\hbar^2 q^2 F_j^2}\right)^{1/3},$$

and $\Delta E_{c1,2}$ are the conduction band offsets between the well and the left and right barriers, respectively.

The tunnel transparency of the structure has been found using the transfer matrix method. Corresponding boundary conditions follow from the integration of the Hamiltonian Eq.(1) across the interface between the left (L) and right (R) regions. For instance, at the interface z=0 the boundary conditions have the form:

$$\Psi_{L\pm}(0) = \Psi_{R\pm}(0) \equiv \Psi_\pm(0),$$

$$\frac{\hbar^2}{2m_z^\pm(L)}\frac{\partial \Psi_{L\pm}}{\partial z} - \frac{\hbar^2}{2m_z^\pm(R)}\frac{\partial \Psi_{R\pm}}{\partial z} \mp k_\parallel \Psi_\pm(0)(\beta_R - \beta_L) = 0. \qquad (5)$$

The transfer matrix has been composed from the set of boundary conditions for all four interfaces likewise Eq.(5).

The vertical charge current through the structure is written below:

$$J = \frac{q}{(2\pi\hbar)^3}\int (f_1 - f_5)\,Tr(t\,t^+)\,dE_z\,d\vec{p}_\|\,, \tag{6}$$

where $f_{1,5}$ are the electron distribution functions in bulk doped *GaN* emitter and collector, respectively. Electron spin splitting in these regions is not implied.

After integration over the directions of the in-plane momentum the current takes the form

$$J = \frac{m_\| q k_B T}{4\pi^2\hbar^3}\int_0^\infty dE_z \int_0^{y_{max}(E_z)} dy\,\Phi(E_z,y)\left[|t_+|^2 + |t_-|^2\right], \tag{7}$$

where

$$\Phi = \frac{\exp(v)-1}{1+\exp(-x-y)+\exp(x+y+v)+\exp(v)},$$

$$x = \frac{E_z - E_f}{k_B T},\quad y = \frac{p_\|^2}{2m_\| k_B T},\quad v = \frac{qV_{ext}}{k_B T}.$$

The limits of the in-plane momentum integration $y_{max}(E_z)$ are determined by the regions where $k_{1,5}$ are real.

Total electron current Eq.(7) can be represented as a sum of electron fluxes in $\pm$ spin states.

Below the spin current is treated as a pseudo-tensor defined in Ref.[12]. The vertical spin current (flux) in z-direction is the axial vector that determines the direction and magnitude of spin polarization:

$$\vec{J}_S = \frac{1}{(2\pi\hbar)^3}\int (f_1 - f_5)\,Tr(t\,\vec{s}\,t^+)\,dE_z\,d\vec{p}_\|\,, \tag{8}$$

where $t = \begin{pmatrix} t_+ & 0 \\ 0 & t_- \end{pmatrix}$, $\vec{s} = \begin{pmatrix} \vec{s}_+ & 0 \\ 0 & \vec{s}_- \end{pmatrix}$, $\vec{s}_\pm = -\vec{s}_\mp = \frac{1}{2}\langle u_\pm|\vec{\sigma}|u_\pm\rangle$. Fig. 2 shows the spin directions $\vec{s}_+$ depending on an electron in-plane momentum.

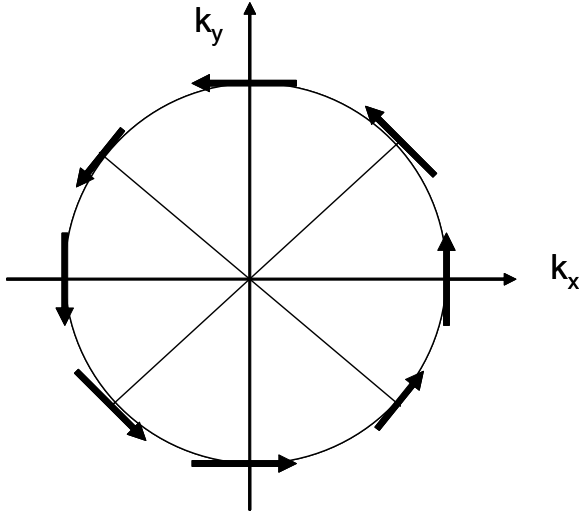

Fig.2. Momentum dependent electron spin vector $\vec{s}_+$.

As long as the distribution functions represent the equilibrium in-plane Fermi distributions $f_{1,5}^0$, the spin current equals zero due to the symmetry of the in-plane electron spectrum and corresponding transmission amplitudes ($t_+(k_\parallel) = t_-(-k_\parallel)$). The lateral electric field $\vec{F}_L$, applied to region 5, breaks the symmetry and induces the non-equilibrium correction to the distribution function $f_5 = f_5^0 + f'$, $f' = \dfrac{q\tau}{m_\parallel} \dfrac{\partial f_5^0}{\partial E} \vec{p}_\parallel \vec{F}_L$ , $\tau$ is the momentum relaxation time. Then, Eq.(8) can be rewritten in the form:

$$\vec{J}_S = -\frac{q\,\tau(2m_\parallel k_B T)^{3/2}}{32\,\pi^2 \hbar^3 m_\parallel} \int_0^\infty dE_z \int_0^{y_{\max}(E_z)} dy\,\sqrt{y}\,\frac{\partial f_5^0}{\partial E}\Big[|t_+|^2 - |t_-|^2\Big]\big(-\vec{x}\,F_{Ly} + \vec{y}\,F_{Lx}\big),$$

(9)

where $\vec{x}, \vec{y}$ are the in-plane unit vectors.

It should be noted that the non-equilibrium correction to the distribution function does not contribute the current-voltage characteristics Eq.(6) as the corresponding term turns zero after an integration over the in-plane momentum directions. Spin current Eq.(9) cannot be represented as a difference of charge currents $J$ calculated separately for electrons in $\pm$ spin states as it was done in Refs.[3,4].

The spin polarization has the direction perpendicular to the lateral field and to a normal to the interface: $\vec{J}_S \sim \left[\vec{n} \times \vec{F}_L\right]$. From the phenomenological standpoint, the effect is similar to the in-plane-current induced spin orientation in an electron gas with the linear-momentum spin splitting [13] or in the Rashba two-dimensional electron gas [14,15]. What discerns the system considered here, from the planar system described in Refs.[14,15], is that the spin polarization appears in the bulk region 5 where the in-plane Rashba spin splitting is not implied: the spin imbalance comes from the vertical tunnelling.

Numerical data have been obtained assuming Dresselhaus coupling constant $\beta_l = 2*10^{-31}\ eV\ m^3$ (Ref.[16]) and the Fermi energy $E_f = 0.1 eV$. Bulk k-linear spin-orbit $\lambda$ and Rashba $\beta_R - \beta_L$ coupling coefficients have been neglected as their role in tunneling is much less than that of the Dresselhaus term that renormalizes the masses of tunneling electrons.

Single-mode tunneling at a particular in-plane momentum $y$ can be characterized with the spin-selective transparency $T_\pm = |t_\pm|^2$ and the tunnel spin polarization $P = \dfrac{T_+ - T_-}{T_+ + T_-}$ shown in Fig.3.

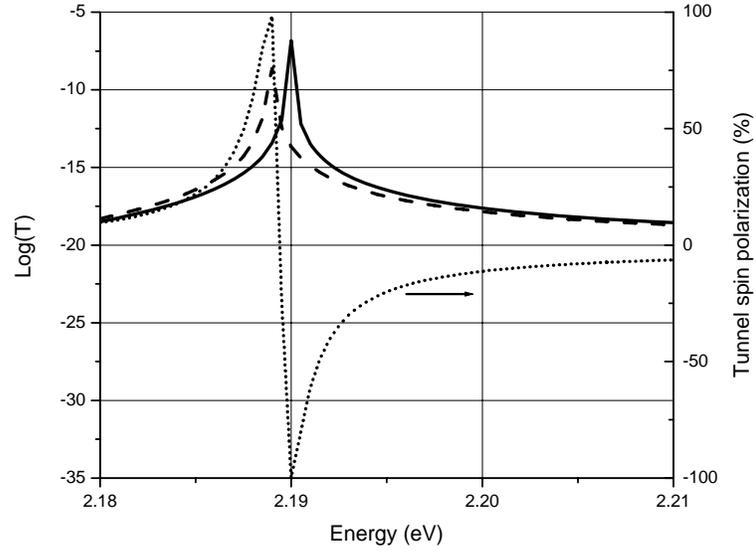

Fig.3. Single-mode transparency and tunnel spin polarization. Continuous line: $T_+$ ; dashed line: $T_-$ ; dotted line: P; $y = 3.86, V_{ext} = -2V$.

As illustrated in Fig.3, the single-mode tunnel spin polarization can be as high as 100% at a resonance, however, it does not mean that the effective spin injection occurs when the tunnel current flows across the structure: the total spin flux contains all possible in-plane modes, weighted with the equilibrium distribution functions, and most important modes are those close in energy to the Fermi level in the collector region 5 to which the lateral electric field is applied.

Figs.4-6 compare the total vertical charge ($J/q$) and spin fluxes across the $30\overset{\circ}{A}/40\overset{\circ}{A}/30\overset{\circ}{A}$ SRTD with Al content $x = y = 1$, $F_L = 2.7*10^5 V/cm$, T=300 K, $\tau = 10^{-13} s$.

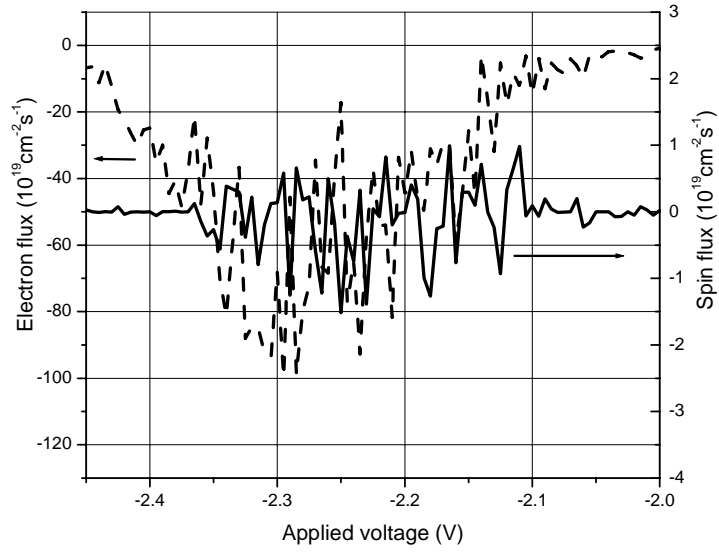

Fig.4. Current-voltage characteristics and spin flux in $30\,\mathring{A}/40\,\mathring{A}/30\,\mathring{A}$ SRTD with Al content $x = y = 1$, $F_L = 2.7*10^5\,V/cm$.

For comparison, Figs. 5 and 6 illustrate the currents through otherwise the same structure of Fig.4, but without polarization fields taken into account. In the example shown in Fig 4, built-in fields make barrier shapes triangular (Zener tunneling) and increase a transparency and a spin flux as compared to the structure where an electrical polarization is not taken into account (Figures 5 and 6).

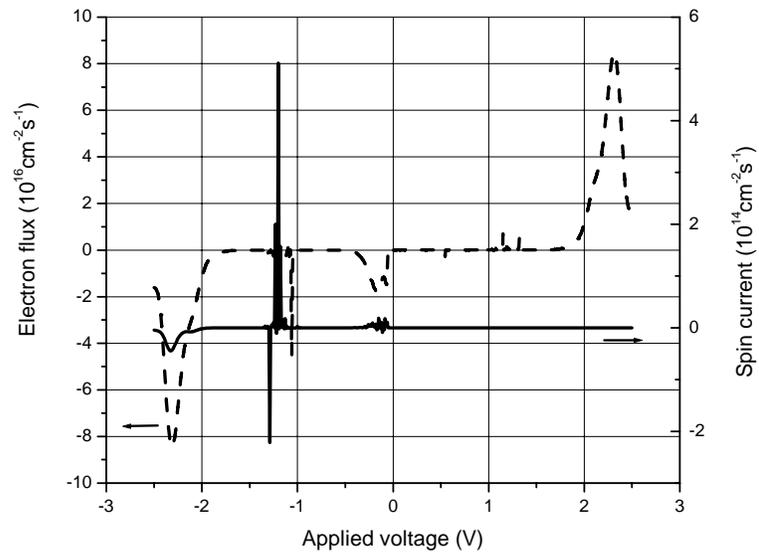

Fig.5. Current-voltage characteristics and spin flux in $30\,\overset{\circ}{A}/40\,\overset{\circ}{A}/30\,\overset{\circ}{A}$ SRTD with Al content $x = y = 1$, $F_L = 2.7*10^5\,V/cm$. Spontaneous and piezoelectric polarizations are absent.

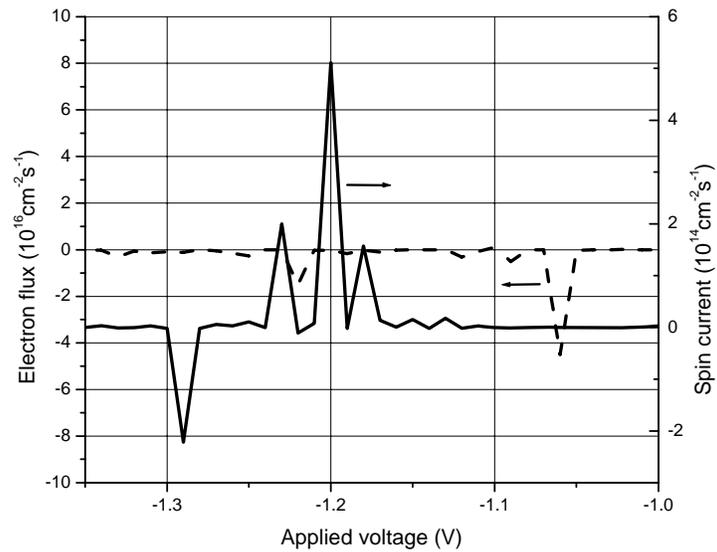

Fig.6. Voltage-scaled data from Fig.5.

Lattice mismatch induced electrical polarization could be the one responsible for the strain-dependent spin polarization observed in Ref.[10]. Piezoelectric fields can be engineered by strain (Al-content in the barriers) that allows manipulating the spin injection efficiency in a perspective spintronic device.

Sheet spin density (spin polarization) in the region 5 can be estimated as $J_S \tau_S$, where $\tau_S$ is the spin relaxation time. Providing the spin relaxation time in GaN-based structures $\tau_S = 0.25\ ps$ [17], from the data shown in Fig.4, it follows that $10^{10} cm^{-2}$ spins are oriented. Extremely long spin relaxation time of $\tau_S = 100\ ps$ in InGaN multiple QW, reported in Ref.[18], may single out GaN/InGaN QW as a perspective structure for an effective spin injector.

References.

1. G. Schmidt, D. Ferrand, L.W. Molenkamp, A.T. Filip, and B.J. van Wees, *Phys. Rev.* B **62** R4790 (2000).

2. E.I.Rashba, Phys. Rev. B62, R16267 (2000).

3. A. Voskoboynikov, S.S. Liu, and C.P. Lee, Phys.Rev. B **58**, 15397 (1998); Phys.Rev. B **59**, 12514 (1999).

4. Voskoboynikov, S.S. Liu, C.P. Lee, and O. Tretyak, J.Appl.Phys, 87, 387 (2000).

5. M.M. Glazov, P.S. Alekseev, M.A. Odnoblyudov, V.M. Chistyakov, S.A. Tarasenko, and I.N. Yassievich, Phys.Rev. B **71**, 155313 (2005).

6. V.I. Litvinov, Phys. Rev. B **68,** 155314 (2003); Appl. Phys. Lett. **89**, 222108

(2006).

7. I. Lo, W.T. Wang, M.H. Gau, S.F. Tsai, and J.C. Chiang, Phys.Rev. B **72**, 245329
 (2005).

8. Ç Kurdak, N. Biyikli, Ü Özgür, H. Morkoç, and V. I. Litvinov, Phys. Rev.B **74**,

113308 (2006).

9. N. Thillosen, Th. Schäpers,a_ N. Kaluza, H. Hardtdegen, and V. A. Guzenko, Appl. Phys.Lett. **88**, 022111 (2006).

10. H. J. Chang, T.W. Chen, J.W. Chen, W. C. Hong, W. C. Tsai, Y. F. Chen, and G.Y. Guo, Phys. Rev.Lett. **98,** 136403 (2007).

11. V.I. Litvinov, A. Manasson, and D. Pavlidis, Appl. Phys. Lett. **85**, 600 (2004).

12. E.I Rashba, Phys. Rev B **70**, 161201(R) (2004).

13. E.L. Ivchenko and G.E. Pikus, JETP Lett. **27**, 604 (1978).

14. L.I. Magarill and M.V. Entin, JETP Lett. **72**, 134 (2000).

15. V.K. Kalevich and V.L. Korenev, JETP Lett. **52**, 230 (1990).

16. H. Cheng, Ç. Kurdak, N. Biyikli, U. Ozgur, H. Morkoç, and V. I. Litvinov, "Spin-Orbit Coupling in AlGaN/AlN/GaN Heterostructures with a Polarization Induced Two-Dimensional Electron Gas", International Conference on Electronic Properties of Two-dimensional Systems, Genova, Italy, July 15 (2007).

17. T. Kuroda, T. Yabushita, T. Kosuge, A. Takeuchi, K. Taniguchi, T. Chinone, and N. Horio, Appl. Phys. Lett. 85, 3116 (2004).

18. M. Jullier, A. Vinatieri, M. Colloci, P. Lefebvre, B. Gil, D. Scalbert, C.A. Tran, R.F. Karlicek, Jr., and J.P. Lascaray, Phys. Stat. Solidi, (b) **216**, 341 (1999).